\documentclass{article}
\usepackage[dblblindworkshop,final,nonatbib]{neurips_2025}

\workshoptitle{Machine Learning and the Physical Sciences}

\usepackage[utf8]{inputenc} %
\usepackage[T1]{fontenc}    %
\usepackage{url}            %
\usepackage{booktabs}       %
\usepackage{amsfonts}       %
\usepackage{nicefrac}       %
\usepackage{microtype}      %
\usepackage[round]{natbib}
\usepackage{xcolor}
\definecolor{linkblue}{HTML}{4e79a7}
\definecolor{linkgreen}{HTML}{59a14f}
\definecolor{linkyellow}{HTML}{edc948}
\definecolor{linkred}{HTML}{e15759}
\definecolor{linkpurple}{HTML}{b07aa1}
\definecolor{linkpink}{HTML}{ff9da7}

\usepackage[
  colorlinks=true,
  linkcolor=linkred,     %
  citecolor=linkblue,    %
  urlcolor=linkgreen,       %
  filecolor=linkpurple,   %
  menucolor=linkyellow,   %
]{hyperref}

\usepackage{amsmath}
\usepackage{graphicx}
\usepackage{bm}

\title{Re-envisioning \textit{Euclid} Galaxy Morphology: Identifying and Interpreting Features with Sparse Autoencoders}

\author{%
  John F. Wu$^{*}$ \\
  Space Telescope Science Institute \\
  Johns Hopkins University \\
  \texttt{jfwu@stsci.edu} \\
  \And
  Michael Walmsley$^{*}$ \\
  University of Toronto \\
  \texttt{m.walmsley@utoronto.ca}
}

\begin{document}

\maketitle

\makeatletter
\def\thefootnote{*}\footnotetext{The authors contributed equally to this work.}\def\thefootnote{\arabic{footnote}}
\makeatother

\begin{abstract}
Sparse Autoencoders (SAEs) can efficiently identify candidate monosemantic features from pretrained neural networks for galaxy morphology. We demonstrate this on Euclid Q1 images using both supervised (Zoobot) and new self-supervised (MAE) models. Our publicly released MAE achieves superhuman image reconstruction performance. While a Principal Component Analysis (PCA) on the supervised model primarily identifies features already aligned with the Galaxy Zoo decision tree, SAEs can identify interpretable features outside of this framework. SAE features also show stronger alignment than PCA with Galaxy Zoo labels. Although challenges in interpretability remain, SAEs provide a powerful engine for discovering astrophysical phenomena beyond the confines of human-defined classification.
\end{abstract}

\section{Introduction}

Galaxy morphology concepts are traditionally encoded in manually-designed taxonomies \citep[e.g.,][]{DeVaucouleurs1963}.
These may miss concepts that are too rare to be manually detected.
New, rare concepts are statistically guaranteed to be represented in the billions of resolved galaxies soon to be imaged by space telescopes like \textit{Euclid} and \textit{Roman} \citep{scaramellaEuclidPreparationEuclid2021,spergel2015widefieldinfrarredsurveytelescopeastrophysics}, so we will need a data-driven method to identify these rare concepts.
Relatedly, the scale of \textit{Euclid} and \textit{Roman} also demands an increasing reliance on deep learning models for interpreting images (e.g. classification, segmentation, feature extraction for multimodal tasks, etc.)
Identifying the concepts learned by our models can inform how we build them and help mitigate the risk of unintended biases or shortcut learning \citep{geirhosShortcutLearningDeep2020}.

Sparse autoencoders (SAEs) are routinely used for interpreting language models by identifying the set of directions in activation space that can together describe any activation vector in a neural network \citep{makhzani2014ksparseautoencoders}. Sparsity encourages these directions to be monosemantic, i.e., corresponding to a single concept \citep{elhage2022toymodelssuperposition,cunninghamSparseAutoencodersFind2023,2025ApJ...980..183W,karvonenSAEBenchComprehensiveBenchmark2025}.
Here, we apply SAEs to reveal the concepts learned by galaxy morphology models.

We construct embeddings for galaxies in the \textit{Euclid} Q1 data release \citep{collaborationEuclidQuickData2025} and then compress those embeddings via SAEs and PCA to identify learned concepts. We show that SAEs efficiently extract candidate morphological features from pretrained networks, comparing against a PCA baseline. For both supervised and self-supervised embeddings, SAEs identify interpretable features aligned with, and beyond the Galaxy Zoo decision trees, while PCA primarily aligns with existing categories.

\section{Representations of Galaxy Morphology}

\textbf{Euclid Imaging and Embeddings}. 
We analyze two types of neural network embeddings for galaxy images in the Euclid Q1 dataset. For supervised embeddings, we use the Zoobot models presented in \cite{euclidcollaborationEuclidQuickData2025b}. These models are pretrained on approximately 1M volunteer-annotated galaxies from four other telescopes and then finetuned using 170k Euclid galaxies with new annotations from Galaxy Zoo Euclid (GZ). Our chosen model, ConvNeXT-Nano, has an encoder with 640-dimensional features.

For self-supervised embeddings, we train a masked autoencoder (MAE, \citealt{heMaskedAutoencodersAre2022}) on 3M Euclid images drawn from the internal Euclid dataset ``RR2''. This is $5-10\times$ larger than recent work applying MAE to astronomy images \citep{fathkouhiAstroMAERedshiftPrediction2024} and applying other self-supervised learning approaches to Euclid images \citep{euclidcollaborationEuclidQuickData2025}.  

Our MAE is a ViT-S\footnote{\url{https://huggingface.co/timm/vit_small_patch16_224.augreg_in21k}} encoder (30.1M parameters, 384-dimensional features) plus a 3-layer decoder, trained for 2 A100-weeks. We make several design changes compared to the standard ConvNeXT/MAE approach in order to achieve stronger reconstruction performance:  

\textit{We use a high masking fraction of 90\%} (vs. 75\% as published). We find that networks of just 1M parameters perform well with conventional masking ratios, suggesting that astronomy images have high redundancy. Work using MAE for video (which is intrinsically high redundancy) shows that higher masking ratios are required to force the network to do more than interpolate \citep{feichtenhoferMaskedAutoencodersSpatiotemporal2022}. 
\begin{figure}
    \centering
    \includegraphics[width=\linewidth]{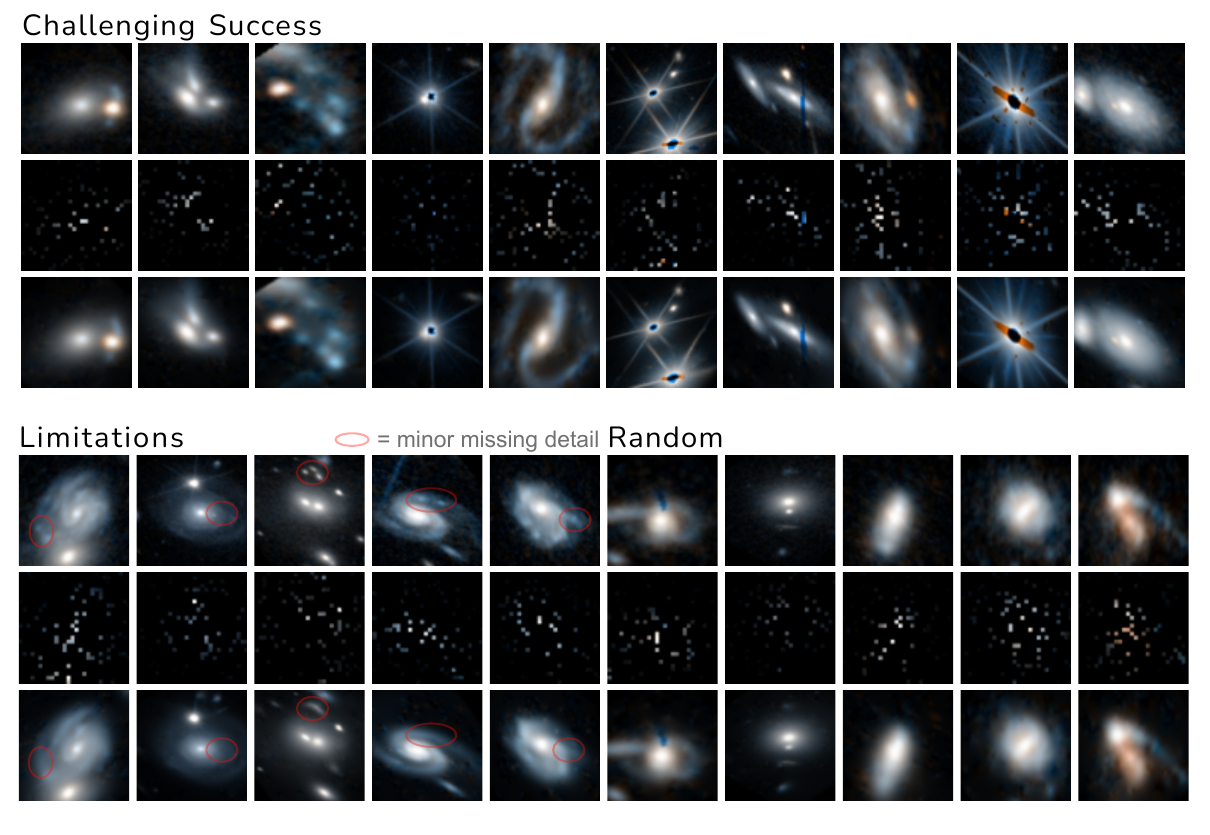}
    \caption{True images (\textit{top row}), masked inputs (\textit{middle row}), and reconstructed outputs (\textit{bottom row}), grouped as follows. Top: challenging images with successful MAE reconstructions. Lower left: images with missing local details (e.g., small starforming clumps, multiple background sources with marginal separation). These cannot be reconstructed if not at least partially included in an unmasked patch. Masking is only applied during training and so these details are still in the embeddings. Lower right: random images (first five) demonstrating that near-perfect reconstructions are the norm. }
    \label{fig:reconstructions}
\end{figure}

\textit{We use a small patch area of 8$\times$8 pixels} (vs. 16$\times$16 as published). Astronomy images are relatively sparse, often with large `blank' areas and compact background/foreground galaxies. Random sampling with large patches misses relevant areas entirely, making prediction impossible. 

\textit{We adjust the frequency of the positional encoding to indicate the angular scale of each image}. This is an astronomical analogy to work using MAE for remote sensing \citep{reedScaleMAEScaleAwareMasked2023}, who adjust the frequency to indicate the ground scale of each satellite image.

Our trained MAE reconstructs masked galaxy features more accurately than professional astronomers. Figure~\ref{fig:reconstructions} shows our MAE infilling 90\% masked galaxy images with near-perfect accuracy. An informal survey of our human colleagues suggests that professional astronomers cannot correctly describe the same masked images. \textit{Superhuman} reconstruction performance indicates that our MAE extracts meaningful features. We also hope that our MAE will be useful for tasks like patch-level anomaly search (find me images that include patches like \textit{these}) and targeted infill (remove \textit{this} small artifact). We release our MAE model on HuggingFace at \url{https://huggingface.co/mwalmsley/euclid-rr2-mae}. We also present a demo of the MAE at \url{https://huggingface.co/spaces/mwalmsley/euclid_masked_autoencoder}.

\begin{figure}[t]
    \centering
    \includegraphics[width=\textwidth]{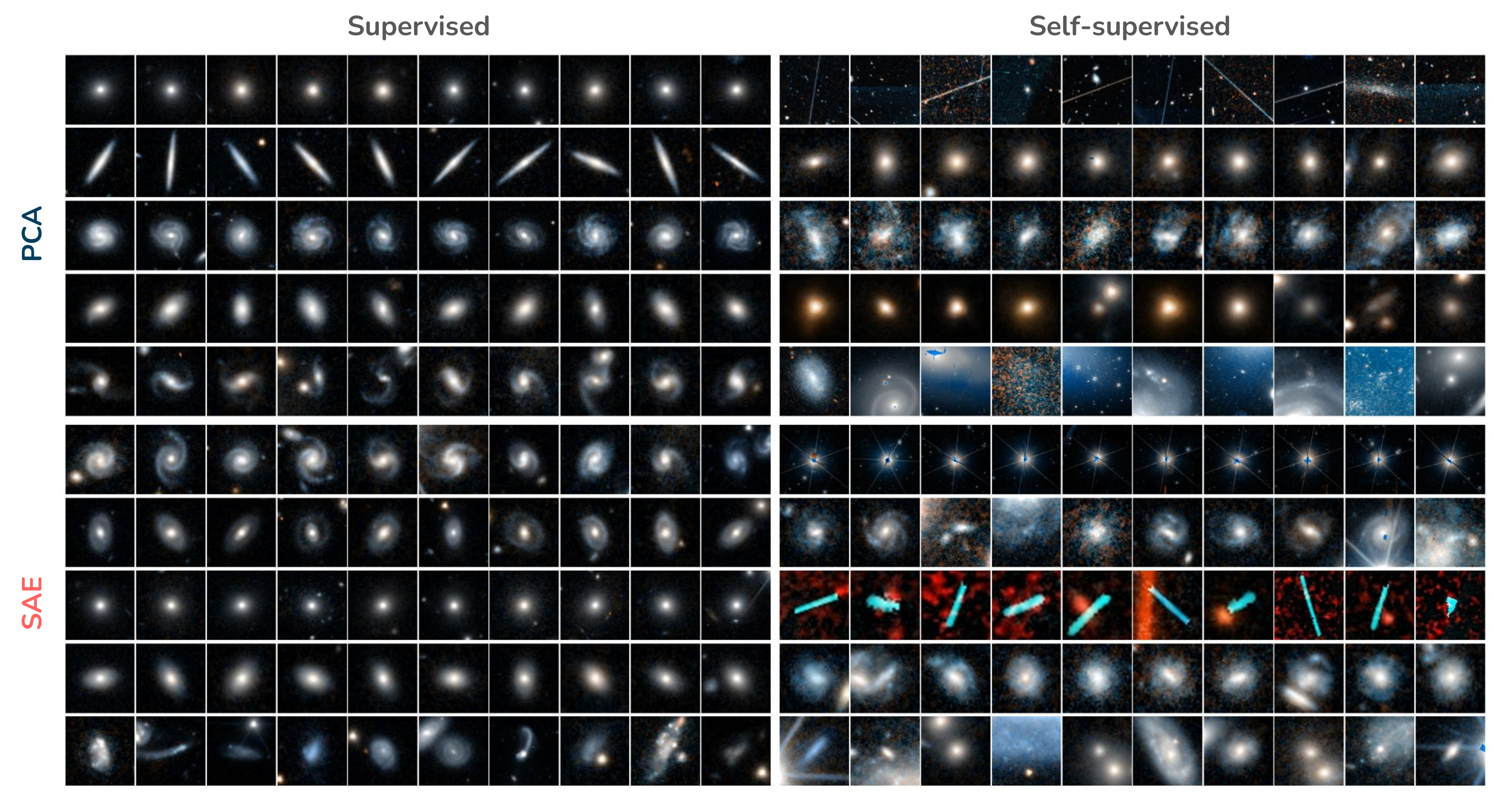}

    \caption{Top 10 examples for first five features extracted via PCA (\textit{upper)}, and SAEs (\textit{lower}), for embeddings extracted from supervised (\textit{left}) and self-supervised (\textit{right}) models.}
    \label{fig:gallery}
\end{figure}

\textbf{Extracting Features from Embeddings}.
SAEs learn sparse representations of the embeddings using an \textit{overcomplete} basis. 
We train a Matryoshka Sparse Autoencoder (SAE; \citep{bussmann2025learningmultilevelfeaturesmatryoshka}) with hierarchical group sizes 64, 64, 128, $\cdots$, 1024, batch top-$k$ sparsity ($k=64$), a small L1 penalty, and auxiliary loss for reviving dead neurons (following \citep{2024arXiv240800657O}). 
For both self-supervised and supervised embeddings, we train for 200 epochs. The code is publicly available at \url{https://github.com/jwuphysics/euclid-galaxy-morphology-saes}.

We use top $k=64$ components in terms of activation frequency. Figure~\ref{fig:gallery} shows the first 5 for each. They explain $\sim 83\%$ of the variance for Zoobot and $\sim 90\%$ of the variance for the MAE).
We also perform PCA decomposition of Zoobot and MAE embeddings; the cumulative explained variance using first 64 components is $\sim 90\%$ for Zoobot and $\sim 95\%$ for the MAE.

\textbf{Interpreting SAE Features via Alignment with Galaxy Zoo.} For each learned feature, we compute Spearman rank correlations ($r$) with all Galaxy Zoo morphology labels and take the maximum absolute correlation.\footnote{See Appendix~\ref{app:predictability} for a similar analysis on whether SAE features are predictable from GZ vote fractions.}  Higher values indicate stronger correlation with known morphological classifications. We report the mean over the top $k$ PCA or SAE features, where ``top'' is defined as the highest rank-ordered eigenvalues for PCA, and the most frequently activated neurons for SAEs.

\section{Results}

\begin{figure}
    \centering
    \includegraphics[width=0.475\textwidth]{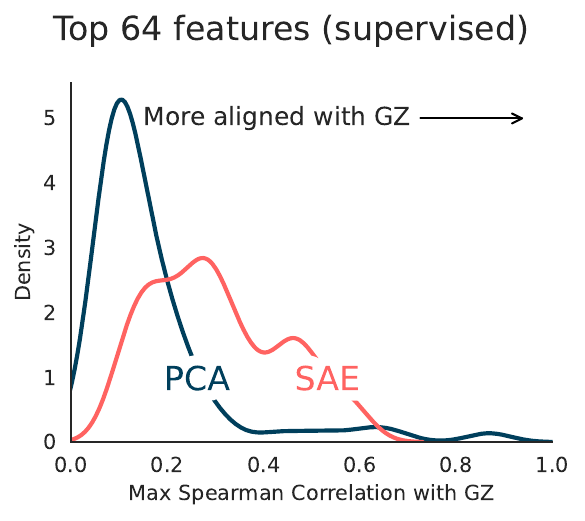}
    \includegraphics[width=0.475\textwidth]{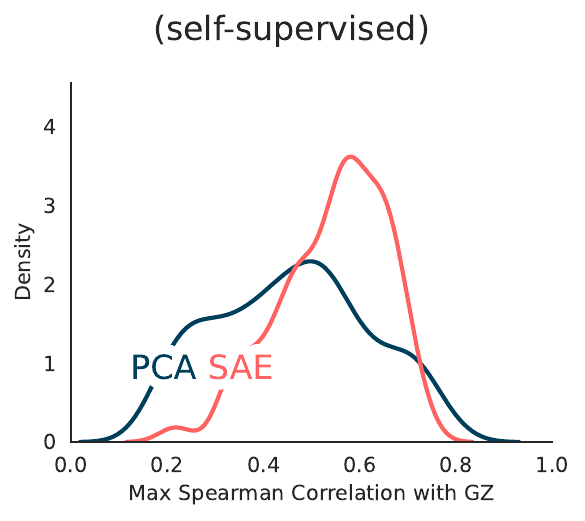}
    \caption{Top SAE features (red) are generally more aligned with GZ morphological classifications than PCA (blue) for both the supervised model embeddings (left) and self-supervised model embeddings (right). Both panels show the distribution of maximum Spearman rank correlation $r$ values between top 64 features and any GZ class. 
    }
    \label{fig:feature_alignment}
\end{figure}

In Figure~\ref{fig:feature_alignment}, we show how well the SAE and PCA features correlate with GZ features. 
An extension of this analysis beyond the top $k=64$ features is presented in the Appendix~\ref{app:beyond} (see, e.g., Figure~\ref{fig:appendix_feature_alignment}).

\textbf{Supervised}:  
The top $k=5$ principal components of the supervised embeddings strongly correlate with GZ features, even more so than the corresponding top 5 SAE features (see Table~\ref{tab:alignment}). When we extend to the top $k=64$, higher-order PCA features show weaker alignment with GZ classes than the SAE features. At the least, this implies that SAE features retain coherency while PCA features become more noisy at $k=64$. We see SAE features outside of the GZ decision tree, e.g., dust lanes in edge-on disk galaxies, elliptical galaxies with bluer companions, etc.

All of the strongest correlations between PCA and SAE features are due to the \texttt{smooth-or-featured-euclid\_*} node in the GZ decision tree. This is unsurprising, as the root node classification routes downstream morphological classifications into ``secondary'' categories. Thus we repeat our alignment tests by omitting this root node, and only using secondary. Overall, we find that alignment with GZ decreases, but we observe the same trends as before.

\textbf{Self-supervised}: The MAE embeddings exhibit surprisingly strong correlations with GZ---they are more correlated than the Zoobot embeddings. We suspect this arises because the MAE embeddings are intent on reconstructing imaging artifacts such as saturated stars or ghosts, which can be diverse and span many pixels. The SAE and PCA features primarily activate on these artifacts, and indeed we see that the most of the strongly aligned GZ classes are \texttt{problem\_*} or \texttt{artifact\_*}.\footnote{Another possibility is that the supervised optimization objective (i.e. aligning to GZ) can (counterintuitively) cram class-separating features into a low-dimensional subspace \cite[e.g.,][]{grigg2021selfsupervisedsupervisedmethodslearn}. 
In other words, the supervised objective can cause embeddings to span a small number of known, dominant morphological classes, whereas the self-supervised case allows representations to disentangle and occupy a broader space of potentially novel features.}

\begin{table}
\caption{Alignment between SAE/PCA Features and Galaxy Zoo vote fractions. We report alignment with ``all'' GZ classes, as well as the alignment with GZ classes beyond the most common classes (``secondary''), for the top $k=5$ and $k=64$ features extracted via PCA and SAEs. Random feature statistics are described in Appendix~\ref{app:random-spearman}. \label{tab:alignment}}
\centering
\footnotesize
\begin{tabular}{lcccc} 
\toprule
Features & \multicolumn{2}{c}{Supervised} & \multicolumn{2}{c}{Self-supervised} \\
 & (all) & (secondary) & (all) & (secondary) \\
\midrule
PCA ($k=5$) & $\bm{0.618 \pm 0.149}$ & $\bm{0.582 \pm 0.097}$ & $0.455 \pm 0.199$ & $0.444 \pm 0.193$ \\
SAE ($k=5$) & $0.402 \pm 0.154$ & $0.371 \pm 0.161$ & $\bm{0.589 \pm 0.067}$ & $\bm{0.589 \pm 0.067}$ \\
\midrule
PCA ($k=64$) & $0.176 \pm 0.138$ & $0.166 \pm 0.137$ & $0.434 \pm 0.174$ & $0.396 \pm 0.178$ \\
SAE ($k=64$) & $\bm{0.296 \pm 0.129}$ & $\bm{0.283 \pm 0.126}$ & $\bm{0.523 \pm 0.123}$ & $\bm{0.509 \pm 0.128}$ \\
\midrule
Random & $0.041 \pm 0.002$ & $0.040 \pm 0.002$ & $0.0072 \pm 0.0003$ & $0.0071 \pm 0.0003$ \\
\bottomrule
\end{tabular}
\end{table}

\section{Discussion}

\textbf{Limitations}.
Unfortunately, some SAE features remain difficult to interpret, despite the benefits of using Matryoshka SAEs to combat feature splitting and absorption \citep{Chanin2024AIF,bussmann2025learningmultilevelfeaturesmatryoshka}.
Still, SAEs surpass our PCA baseline approach. PCA features become unidentifiable after $\sim 30$ features and show no alignment with GZ. In contrast, our SAE delivers an order of magnitude more aligned features (see Appendix~\ref{app:beyond} for more details).

\textbf{Scientific Discovery}.
SAEs can extract interpretable features from both supervised and self-supervised models at scale \citep{gao2024scalingevaluatingsparseautoencoders}. Sparse activations are  steerable and can be readily interpreted through automated means \citep{cunninghamSparseAutoencodersFind2023,2024arXiv240800657O,ye2024steering}.
Although there are concerns about whether SAEs are effective at learning \textit{known} concepts, they excel at identifying \textit{new} features compared to other methods \citep{muhamed2024decodingdarkmatterspecialized,movva2025sparseautoencodershypothesisgeneration,marks2025auditinglanguagemodelshidden,peng2025usesparseautoencodersdiscover}. Thus, SAEs have tremendous potential to aid scientific discovery by categorizing and sifting through next-generation astronomical datasets (see also, e.g., \citealt{2021A&C....3600481L,oryan2025identifyingastrophysicalanomalies996}). By extracting features from \textit{self-supervised} embedding models, SAEs can act as scalable discovery engines that surface rare or anomalous astrophysical phenomena for future targeted investigation.

\bibliographystyle{plainnat}
\bibliography{main}

\begin{thebibliography}{28}
\providecommand{\natexlab}[1]{#1}
\providecommand{\url}[1]{\texttt{#1}}
\expandafter\ifx\csname urlstyle\endcsname\relax
  \providecommand{\doi}[1]{doi: #1}\else
  \providecommand{\doi}{doi: \begingroup \urlstyle{rm}\Url}\fi

\bibitem[Bussmann et~al.(2025)Bussmann, Nabeshima, Karvonen, and Nanda]{bussmann2025learningmultilevelfeaturesmatryoshka}
Bart Bussmann, Noa Nabeshima, Adam Karvonen, and Neel Nanda.
\newblock Learning multi-level features with matryoshka sparse autoencoders, 2025.
\newblock URL \url{https://arxiv.org/abs/2503.17547}.

\bibitem[Chanin et~al.(2024)Chanin, Wilken-Smith, Dulka, Bhatnagar, and Bloom]{Chanin2024AIF}
David Chanin, James Wilken-Smith, Tom'avs Dulka, Hardik Bhatnagar, and Joseph Bloom.
\newblock A is for absorption: Studying feature splitting and absorption in sparse autoencoders.
\newblock \emph{ArXiv}, abs/2409.14507, 2024.
\newblock URL \url{https://api.semanticscholar.org/CorpusID:272827216}.

\bibitem[Cunningham et~al.(2023)Cunningham, Ewart, Riggs, Huben, and Sharkey]{cunninghamSparseAutoencodersFind2023}
Hoagy Cunningham, Aidan Ewart, Logan Riggs, Robert Huben, and Lee Sharkey.
\newblock Sparse {Autoencoders} {Find} {Highly} {Interpretable} {Features} in {Language} {Models}, October 2023.
\newblock URL \url{http://arxiv.org/abs/2309.08600}.
\newblock arXiv:2309.08600 [cs].

\bibitem[De~Vaucouleurs(1963)]{DeVaucouleurs1963}
Gerard De~Vaucouleurs.
\newblock Revised {Classification} of 1500 {Bright} {Galaxies}.
\newblock \emph{The Astrophysical Journal Supplement Series}, 8:\penalty0 31, 1963.

\bibitem[Elhage et~al.(2022)Elhage, Hume, Olsson, Schiefer, Henighan, Kravec, Hatfield-Dodds, Lasenby, Drain, Chen, Grosse, McCandlish, Kaplan, Amodei, Wattenberg, and Olah]{elhage2022toymodelssuperposition}
Nelson Elhage, Tristan Hume, Catherine Olsson, Nicholas Schiefer, Tom Henighan, Shauna Kravec, Zac Hatfield-Dodds, Robert Lasenby, Dawn Drain, Carol Chen, Roger Grosse, Sam McCandlish, Jared Kaplan, Dario Amodei, Martin Wattenberg, and Christopher Olah.
\newblock Toy models of superposition, 2022.
\newblock URL \url{https://arxiv.org/abs/2209.10652}.

\bibitem[{Euclid Collaboration} et~al.(2025{\natexlab{a}}){Euclid Collaboration}, Aussel, Tereno, Schirmer, Alguero, Altieri, Balbinot, Boer, Casenove, Corcho-Caballero, Furusawa, Furusawa, Hudson, Jahnke, Libet, Macias-Perez, Masoumzadeh, Mohr, Odier, Scott, Vassallo, Kleijn, Zacchei, Aghanim, Amara, Andreon, Auricchio, Awan, Azzollini, Baccigalupi, Baldi, Balestra, Bardelli, Basset, Battaglia, Belikov, Bender, Biviano, Bonchi, Bonino, Branchini, Brescia, Brinchmann, Camera, Cañas-Herrera, Capobianco, Carbone, Cardone, Carretero, Casas, Castander, Castellano, Castignani, Cavuoti, Chambers, Cimatti, Colodro-Conde, Congedo, Conselice, Conversi, Copin, Courbin, Courtois, Cropper, Cuby, Silva, Silva, Degaudenzi, Jong, Lucia, Giorgio, Dinis, Dolding, Dole, Douspis, Dubath, Duncan, Dupac, Dusini, Ealet, Escoffier, Fabricius, Farina, Farinelli, Faustini, Ferriol, Fotopoulou, Fourmanoit, Frailis, Franceschi, Franzetti, Galeotta, George, Gillard, Gillis, Giocoli, Gómez-Alvarez, Gracia-Carpio, Granett, Grazian,
  Grupp, Guzzo, Gwyn, Haugan, Herent, Hoar, Hoekstra, Holliman, Holmes, Hook, Hormuth, Hornstrup, Hudelot, Ilić, Jhabvala, Joachimi, Keihänen, Kermiche, Kiessling, Kubik, Kuijken, Kümmel, Kunz, Kurki-Suonio, Lahav, Boulc'h, Brun, Mignant, Liebing, Ligori, Lilje, Lindholm, Lloro, Mainetti, Maino, Maiorano, Mansutti, Marcin, Marggraf, Markovic, Martinelli, Martinet, Marulli, Massey, Maurogordato, McCracken, Medinaceli, Mei, Melchior, Mellier, Meneghetti, Merlin, Meylan, Mora, Moresco, Morris, Moscardini, Mourre, Nakajima, Neissner, Nichol, Niemi, Nightingale, Nutma, Padilla, Paltani, Pasian, Peacock, Pedersen, Percival, Pettorino, Pires, Polenta, Pollack, Poncet, Popa, Pozzetti, Racca, Raison, Rebolo, Renzi, Rhodes, Riccio, Rix, Romelli, Roncarelli, Rossetti, Rusholme, Saglia, Sakr, Sánchez, Sapone, Sartoris, Sauvage, Schewtschenko, Schneider, Scodeggio, Secroun, Sefusatti, Seidel, Seiffert, Serrano, Simon, Sirignano, Sirri, Skottfelt, Mancini, Stanco, Steinwagner, Surace, Tallada-Crespí, Tavagnacco,
  Taylor, Teplitz, Tessore, Toft, Toledo-Moreo, Torradeflot, Tsyganov, Tutusaus, Valentijn, Valenziano, Valiviita, Veropalumbo, Wang, Weller, Williams, Zamorani, Zerbi, Zucca, Allevato, Ballardini, Blake, Bolzonella, Bozzo, Burigana, Cabanac, Calabrese, Cappi, Ferdinando, Vigo, Gabarra, Hartley, Huertas-Company, Martín-Fleitas, Matthew, Maturi, Mauri, Metcalf, Pezzotta, Pöntinen, Porciani, Risso, Scottez, Sereno, Tenti, Viel, Wiesmann, Akrami, Alvi, Andika, Anselmi, Archidiacono, Atrio-Barandela, Avila, Bergamini, Bertacca, Bethermin, Bisigello, Blanchard, Blot, Böhringer, Borgani, Borlaff, Brown, Bruton, Buitrago, Calabro, Calderone, Quevedo, Caro, Carvalho, Castro, Charles, Cogato, Conseil, Cooray, Costanzi, Cucciati, Davini, Paolis, Desprez, Díaz-Sánchez, Diaz, Domizio, Diego, Dimauro, Duc, Enia, Fang, Ferguson, Ferrari, Finoguenov, Fontana, Fontanot, Franco, García-Bellido, Gasparetto, Gavazzi, Gaztanaga, Giacomini, Gianotti, Gonzalez, Gozaliasl, Gruppuso, Guidi, Gutierrez, Hall,
  Hernández-Monteagudo, Hildebrandt, Hjorth, Jacobson, Joudaki, Kajava, Kang, Kansal, Karagiannis, Kiiveri, Kirkpatrick, Kruk, Lacasa, Laigle, Lattanzi, Brun, Graet, Legrand, Lembo, Lepori, Leroy, Lesci, Lesgourgues, Leuzzi, Liaudat, Loureiro, Magliocchetti, Magnier, Mancini, Mannucci, Maoli, Martins, Maurin, McPartland, Melin, Migliaccio, Miluzio, Monaco, Montoro, Moretti, Morgante, Murray, Nadathur, Naidoo, Navarro-Alsina, Nesseris, Nicastro, Oguri, Passalacqua, Paterson, Patrizii, Pisani, Potter, Quai, Radovich, Reimberg, Rocci, Rodighiero, Rollins, Sacquegna, Sahlén, Sanders, Sarpa, Scarlata, Schaye, Schneider, Schultheis, Sciotti, Scognamiglio, Sellentin, Shankar, Smith, Soubrie, Stanford, Tanidis, Tao, Testera, Tewes, Teyssier, Tosi, Troja, Tucci, Valieri, Venhola, Vergani, Vernizzi, Verza, Vielzeuf, Walton, Weaver, Wilde, and Zalesky]{collaborationEuclidQuickData2025}
{}~{Euclid Collaboration}, H.~Aussel, I.~Tereno, M.~Schirmer, G.~Alguero, B.~Altieri, E.~Balbinot, T.~de Boer, P.~Casenove, P.~Corcho-Caballero, H.~Furusawa, J.~Furusawa, M.~J. Hudson, K.~Jahnke, G.~Libet, J.~Macias-Perez, N.~Masoumzadeh, J.~J. Mohr, J.~Odier, D.~Scott, T.~Vassallo, G.~Verdoes Kleijn, A.~Zacchei, N.~Aghanim, A.~Amara, S.~Andreon, N.~Auricchio, S.~Awan, R.~Azzollini, C.~Baccigalupi, M.~Baldi, A.~Balestra, S.~Bardelli, A.~Basset, P.~Battaglia, A.~N. Belikov, R.~Bender, A.~Biviano, A.~Bonchi, D.~Bonino, E.~Branchini, M.~Brescia, J.~Brinchmann, S.~Camera, G.~Cañas-Herrera, V.~Capobianco, C.~Carbone, V.~F. Cardone, J.~Carretero, S.~Casas, F.~J. Castander, M.~Castellano, G.~Castignani, S.~Cavuoti, K.~C. Chambers, A.~Cimatti, C.~Colodro-Conde, G.~Congedo, C.~J. Conselice, L.~Conversi, Y.~Copin, F.~Courbin, H.~M. Courtois, M.~Cropper, J.-G. Cuby, A.~Da Silva, R.~da Silva, H.~Degaudenzi, J.~T. A.~de Jong, G.~De Lucia, A.~M.~Di Giorgio, J.~Dinis, C.~Dolding, H.~Dole, M.~Douspis, F.~Dubath, C.~A.~J.
  Duncan, X.~Dupac, S.~Dusini, A.~Ealet, S.~Escoffier, M.~Fabricius, M.~Farina, R.~Farinelli, F.~Faustini, S.~Ferriol, S.~Fotopoulou, N.~Fourmanoit, M.~Frailis, E.~Franceschi, P.~Franzetti, S.~Galeotta, K.~George, W.~Gillard, B.~Gillis, C.~Giocoli, P.~Gómez-Alvarez, J.~Gracia-Carpio, B.~R. Granett, A.~Grazian, F.~Grupp, L.~Guzzo, S.~Gwyn, S.~V.~H. Haugan, O.~Herent, J.~Hoar, H.~Hoekstra, M.~S. Holliman, W.~Holmes, I.~M. Hook, F.~Hormuth, A.~Hornstrup, P.~Hudelot, S.~Ilić, M.~Jhabvala, B.~Joachimi, E.~Keihänen, S.~Kermiche, A.~Kiessling, B.~Kubik, K.~Kuijken, M.~Kümmel, M.~Kunz, H.~Kurki-Suonio, O.~Lahav, Q.~Le Boulc'h, A.~M. C.~Le Brun, D.~Le Mignant, P.~Liebing, S.~Ligori, P.~B. Lilje, V.~Lindholm, I.~Lloro, G.~Mainetti, D.~Maino, E.~Maiorano, O.~Mansutti, S.~Marcin, O.~Marggraf, K.~Markovic, M.~Martinelli, N.~Martinet, F.~Marulli, R.~Massey, S.~Maurogordato, H.~J. McCracken, E.~Medinaceli, S.~Mei, M.~Melchior, Y.~Mellier, M.~Meneghetti, E.~Merlin, G.~Meylan, A.~Mora, M.~Moresco, P.~W. Morris,
  L.~Moscardini, S.~Mourre, R.~Nakajima, C.~Neissner, R.~C. Nichol, S.-M. Niemi, J.~W. Nightingale, T.~Nutma, C.~Padilla, S.~Paltani, F.~Pasian, J.~A. Peacock, K.~Pedersen, W.~J. Percival, V.~Pettorino, S.~Pires, G.~Polenta, J.~E. Pollack, M.~Poncet, L.~A. Popa, L.~Pozzetti, G.~D. Racca, F.~Raison, R.~Rebolo, A.~Renzi, J.~Rhodes, G.~Riccio, H.-W. Rix, E.~Romelli, M.~Roncarelli, E.~Rossetti, B.~Rusholme, R.~Saglia, Z.~Sakr, A.~G. Sánchez, D.~Sapone, B.~Sartoris, M.~Sauvage, J.~A. Schewtschenko, P.~Schneider, M.~Scodeggio, A.~Secroun, E.~Sefusatti, G.~Seidel, M.~Seiffert, S.~Serrano, P.~Simon, C.~Sirignano, G.~Sirri, J.~Skottfelt, A.~Spurio Mancini, L.~Stanco, J.~Steinwagner, C.~Surace, P.~Tallada-Crespí, D.~Tavagnacco, A.~N. Taylor, H.~I. Teplitz, N.~Tessore, S.~Toft, R.~Toledo-Moreo, F.~Torradeflot, A.~Tsyganov, I.~Tutusaus, E.~A. Valentijn, L.~Valenziano, J.~Valiviita, A.~Veropalumbo, Y.~Wang, J.~Weller, O.~R. Williams, G.~Zamorani, F.~M. Zerbi, E.~Zucca, V.~Allevato, M.~Ballardini, R.~P. Blake,
  M.~Bolzonella, E.~Bozzo, C.~Burigana, R.~Cabanac, M.~Calabrese, A.~Cappi, D.~Di Ferdinando, J.~A.~Escartin Vigo, L.~Gabarra, W.~G. Hartley, M.~Huertas-Company, J.~Martín-Fleitas, S.~Matthew, M.~Maturi, N.~Mauri, R.~B. Metcalf, A.~Pezzotta, M.~Pöntinen, C.~Porciani, I.~Risso, V.~Scottez, M.~Sereno, M.~Tenti, M.~Viel, M.~Wiesmann, Y.~Akrami, S.~Alvi, I.~T. Andika, S.~Anselmi, M.~Archidiacono, F.~Atrio-Barandela, S.~Avila, P.~Bergamini, D.~Bertacca, M.~Bethermin, L.~Bisigello, A.~Blanchard, L.~Blot, H.~Böhringer, S.~Borgani, A.~S. Borlaff, M.~L. Brown, S.~Bruton, F.~Buitrago, A.~Calabro, G.~Calderone, B.~Camacho Quevedo, F.~Caro, C.~S. Carvalho, T.~Castro, Y.~Charles, F.~Cogato, S.~Conseil, A.~R. Cooray, M.~Costanzi, O.~Cucciati, S.~Davini, F.~De Paolis, G.~Desprez, A.~Díaz-Sánchez, J.~J. Diaz, S.~Di Domizio, J.~M. Diego, P.~Dimauro, P.-A. Duc, A.~Enia, Y.~Fang, A.~M.~N. Ferguson, A.~G. Ferrari, A.~Finoguenov, A.~Fontana, F.~Fontanot, A.~Franco, J.~García-Bellido, T.~Gasparetto, R.~Gavazzi, E.~Gaztanaga,
  F.~Giacomini, F.~Gianotti, A.~H. Gonzalez, G.~Gozaliasl, A.~Gruppuso, M.~Guidi, C.~M. Gutierrez, A.~Hall, C.~Hernández-Monteagudo, H.~Hildebrandt, J.~Hjorth, J.~Jacobson, S.~Joudaki, J.~J.~E. Kajava, Y.~Kang, V.~Kansal, D.~Karagiannis, K.~Kiiveri, C.~C. Kirkpatrick, S.~Kruk, F.~Lacasa, C.~Laigle, M.~Lattanzi, V.~Le Brun, J.~Le Graet, L.~Legrand, M.~Lembo, F.~Lepori, G.~Leroy, G.~F. Lesci, J.~Lesgourgues, L.~Leuzzi, T.~I. Liaudat, A.~Loureiro, M.~Magliocchetti, E.~A. Magnier, C.~Mancini, F.~Mannucci, R.~Maoli, C.~J. A.~P. Martins, L.~Maurin, C.~J.~R. McPartland, J.-B. Melin, M.~Migliaccio, M.~Miluzio, P.~Monaco, A.~Montoro, C.~Moretti, G.~Morgante, C.~Murray, S.~Nadathur, K.~Naidoo, A.~Navarro-Alsina, S.~Nesseris, L.~Nicastro, M.~Oguri, F.~Passalacqua, K.~Paterson, L.~Patrizii, A.~Pisani, D.~Potter, S.~Quai, M.~Radovich, P.~Reimberg, P.-F. Rocci, G.~Rodighiero, R.~P. Rollins, S.~Sacquegna, M.~Sahlén, D.~B. Sanders, E.~Sarpa, C.~Scarlata, J.~Schaye, A.~Schneider, M.~Schultheis, D.~Sciotti, D.~Scognamiglio,
  E.~Sellentin, F.~Shankar, L.~C. Smith, E.~Soubrie, S.~A. Stanford, K.~Tanidis, C.~Tao, G.~Testera, M.~Tewes, R.~Teyssier, S.~Tosi, A.~Troja, M.~Tucci, C.~Valieri, A.~Venhola, D.~Vergani, F.~Vernizzi, G.~Verza, P.~Vielzeuf, N.~A. Walton, J.~R. Weaver, J.~Wilde, and L.~Zalesky.
\newblock Euclid {Quick} {Data} {Release} ({Q1}) -- {Data} release overview, March 2025{\natexlab{a}}.
\newblock URL \url{http://arxiv.org/abs/2503.15302}.
\newblock arXiv:2503.15302 [astro-ph].

\bibitem[{Euclid Collaboration} et~al.(2025{\natexlab{b}}){Euclid Collaboration}, Siudek, Huertas-Company, Smith, Martinez-Solaeche, Lanusse, Ho, Angeloudi, Cunha, Sánchez, Dunn, Fu, Iglesias-Navarro, Junais, Knapen, Laloux, Mezcua, Roster, Stevens, Vega-Ferrero, Aghanim, Altieri, Amara, Andreon, Auricchio, Aussel, Baccigalupi, Baldi, Bardelli, Battaglia, Biviano, Bonchi, Branchini, Brescia, Brinchmann, Camera, Cañas-Herrera, Capobianco, Carbone, Carretero, Casas, Castander, Castellano, Castignani, Cavuoti, Chambers, Cimatti, Colodro-Conde, Congedo, Conselice, Conversi, Copin, Courbin, Courtois, Cropper, Silva, Degaudenzi, Lucia, Giorgio, Dinis, Dolding, Dole, Dubath, Duncan, Dupac, Dusini, Escoffier, Farina, Farinelli, Faustini, Ferriol, Finelli, Fotopoulou, Frailis, Franceschi, Galeotta, George, Gillis, Giocoli, Gracia-Carpio, Granett, Grazian, Grupp, Gwyn, Haugan, Holmes, Hook, Hormuth, Hornstrup, Jahnke, Jhabvala, Keihänen, Kermiche, Kiessling, Kubik, Kümmel, Kunz, Kurki-Suonio, Boulc'h, Brun,
  Mignant, Ligori, Lilje, Lindholm, Lloro, Mainetti, Maino, Maiorano, Mansutti, Marcin, Marggraf, Martinelli, Martinet, Marulli, Massey, Maurogordato, McCracken, Medinaceli, Mei, Melchior, Mellier, Meneghetti, Merlin, Meylan, Mora, Moresco, Moscardini, Nakajima, Neissner, Niemi, Nightingale, Padilla, Paltani, Pasian, Pedersen, Percival, Pettorino, Pires, Polenta, Poncet, Popa, Pozzetti, Raison, Renzi, Rhodes, Riccio, Romelli, Roncarelli, Saglia, Sakr, Sánchez, Sapone, Sartoris, Schewtschenko, Schneider, Schrabback, Scodeggio, Secroun, Seidel, Seiffert, Serrano, Simon, Sirignano, Sirri, Stanco, Steinwagner, Tallada-Crespí, Taylor, Tereno, Toft, Toledo-Moreo, Torradeflot, Tutusaus, Valenziano, Valiviita, Vassallo, Kleijn, Veropalumbo, Wang, Weller, Zacchei, Zamorani, Zerbi, Zinchenko, Zucca, Allevato, Ballardini, Bolzonella, Bozzo, Burigana, Cabanac, Cappi, Ferdinando, Vigo, Gabarra, Martín-Fleitas, Matthew, Mauri, Metcalf, Pezzotta, Pöntinen, Porciani, Risso, Scottez, Sereno, Tenti, Viel, Wiesmann, Akrami,
  Andika, Anselmi, Archidiacono, Atrio-Barandela, Benoist, Benson, Bertacca, Bethermin, Bisigello, Blanchard, Blot, Brown, Bruton, Calabro, Quevedo, Caro, Carvalho, Castro, Charles, Cogato, Cooray, Cucciati, Davini, Paolis, Desprez, Díaz-Sánchez, Diaz, Domizio, Diego, Duc, Enia, Fang, Ferrari, Ferreira, Finoguenov, Fontana, Franco, Ganga, García-Bellido, Gasparetto, Gautard, Gaztanaga, Giacomini, Gianotti, Gozaliasl, Guidi, Gutierrez, Hall, Hartley, Hemmati, Hernández-Monteagudo, Hildebrandt, Hjorth, Kajava, Kang, Kansal, Karagiannis, Kiiveri, Kirkpatrick, Kruk, Graet, Legrand, Lembo, Lepori, Leroy, Lesci, Lesgourgues, Leuzzi, Liaudat, Loureiro, Macias-Perez, Maggio, Magliocchetti, Magnier, Mannucci, Maoli, Martins, Maurin, Miluzio, Monaco, Moretti, Morgante, Murray, Naidoo, Navarro-Alsina, Nesseris, Passalacqua, Paterson, Patrizii, Pisani, Potter, Quai, Radovich, Sacquegna, Sahlén, Sanders, Sarpa, Schneider, Sciotti, Scognamiglio, Sellentin, Smith, Tanidis, Testera, Teyssier, Tosi, Troja, Tucci,
  Valieri, Venhola, Vergani, Verza, Vielzeuf, Walton, and Sorce]{euclidcollaborationEuclidQuickData2025}
{}~{Euclid Collaboration}, M.~Siudek, M.~Huertas-Company, M.~Smith, G.~Martinez-Solaeche, F.~Lanusse, S.~Ho, E.~Angeloudi, P.~A.~C. Cunha, H.~Domínguez Sánchez, M.~Dunn, Y.~Fu, P.~Iglesias-Navarro, J.~Junais, J.~H. Knapen, B.~Laloux, M.~Mezcua, W.~Roster, G.~Stevens, J.~Vega-Ferrero, N.~Aghanim, B.~Altieri, A.~Amara, S.~Andreon, N.~Auricchio, H.~Aussel, C.~Baccigalupi, M.~Baldi, S.~Bardelli, P.~Battaglia, A.~Biviano, A.~Bonchi, E.~Branchini, M.~Brescia, J.~Brinchmann, S.~Camera, G.~Cañas-Herrera, V.~Capobianco, C.~Carbone, J.~Carretero, S.~Casas, F.~J. Castander, M.~Castellano, G.~Castignani, S.~Cavuoti, K.~C. Chambers, A.~Cimatti, C.~Colodro-Conde, G.~Congedo, C.~J. Conselice, L.~Conversi, Y.~Copin, F.~Courbin, H.~M. Courtois, M.~Cropper, A.~Da Silva, H.~Degaudenzi, G.~De Lucia, A.~M.~Di Giorgio, J.~Dinis, C.~Dolding, H.~Dole, F.~Dubath, C.~A.~J. Duncan, X.~Dupac, S.~Dusini, S.~Escoffier, M.~Farina, R.~Farinelli, F.~Faustini, S.~Ferriol, F.~Finelli, S.~Fotopoulou, M.~Frailis, E.~Franceschi, S.~Galeotta,
  K.~George, B.~Gillis, C.~Giocoli, J.~Gracia-Carpio, B.~R. Granett, A.~Grazian, F.~Grupp, S.~Gwyn, S.~V.~H. Haugan, W.~Holmes, I.~M. Hook, F.~Hormuth, A.~Hornstrup, K.~Jahnke, M.~Jhabvala, E.~Keihänen, S.~Kermiche, A.~Kiessling, B.~Kubik, M.~Kümmel, M.~Kunz, H.~Kurki-Suonio, Q.~Le Boulc'h, A.~M. C.~Le Brun, D.~Le Mignant, S.~Ligori, P.~B. Lilje, V.~Lindholm, I.~Lloro, G.~Mainetti, D.~Maino, E.~Maiorano, O.~Mansutti, S.~Marcin, O.~Marggraf, M.~Martinelli, N.~Martinet, F.~Marulli, R.~Massey, S.~Maurogordato, H.~J. McCracken, E.~Medinaceli, S.~Mei, M.~Melchior, Y.~Mellier, M.~Meneghetti, E.~Merlin, G.~Meylan, A.~Mora, M.~Moresco, L.~Moscardini, R.~Nakajima, C.~Neissner, S.-M. Niemi, J.~W. Nightingale, C.~Padilla, S.~Paltani, F.~Pasian, K.~Pedersen, W.~J. Percival, V.~Pettorino, S.~Pires, G.~Polenta, M.~Poncet, L.~A. Popa, L.~Pozzetti, F.~Raison, A.~Renzi, J.~Rhodes, G.~Riccio, E.~Romelli, M.~Roncarelli, R.~Saglia, Z.~Sakr, A.~G. Sánchez, D.~Sapone, B.~Sartoris, J.~A. Schewtschenko, P.~Schneider,
  T.~Schrabback, M.~Scodeggio, A.~Secroun, G.~Seidel, M.~Seiffert, S.~Serrano, P.~Simon, C.~Sirignano, G.~Sirri, L.~Stanco, J.~Steinwagner, P.~Tallada-Crespí, A.~N. Taylor, I.~Tereno, S.~Toft, R.~Toledo-Moreo, F.~Torradeflot, I.~Tutusaus, L.~Valenziano, J.~Valiviita, T.~Vassallo, G.~Verdoes Kleijn, A.~Veropalumbo, Y.~Wang, J.~Weller, A.~Zacchei, G.~Zamorani, F.~M. Zerbi, I.~A. Zinchenko, E.~Zucca, V.~Allevato, M.~Ballardini, M.~Bolzonella, E.~Bozzo, C.~Burigana, R.~Cabanac, A.~Cappi, D.~Di Ferdinando, J.~A.~Escartin Vigo, L.~Gabarra, J.~Martín-Fleitas, S.~Matthew, N.~Mauri, R.~B. Metcalf, A.~Pezzotta, M.~Pöntinen, C.~Porciani, I.~Risso, V.~Scottez, M.~Sereno, M.~Tenti, M.~Viel, M.~Wiesmann, Y.~Akrami, I.~T. Andika, S.~Anselmi, M.~Archidiacono, F.~Atrio-Barandela, C.~Benoist, K.~Benson, D.~Bertacca, M.~Bethermin, L.~Bisigello, A.~Blanchard, L.~Blot, M.~L. Brown, S.~Bruton, A.~Calabro, B.~Camacho Quevedo, F.~Caro, C.~S. Carvalho, T.~Castro, Y.~Charles, F.~Cogato, A.~R. Cooray, O.~Cucciati, S.~Davini, F.~De
  Paolis, G.~Desprez, A.~Díaz-Sánchez, J.~J. Diaz, S.~Di Domizio, J.~M. Diego, P.-A. Duc, A.~Enia, Y.~Fang, A.~G. Ferrari, P.~G. Ferreira, A.~Finoguenov, A.~Fontana, A.~Franco, K.~Ganga, J.~García-Bellido, T.~Gasparetto, V.~Gautard, E.~Gaztanaga, F.~Giacomini, F.~Gianotti, G.~Gozaliasl, M.~Guidi, C.~M. Gutierrez, A.~Hall, W.~G. Hartley, S.~Hemmati, C.~Hernández-Monteagudo, H.~Hildebrandt, J.~Hjorth, J.~J.~E. Kajava, Y.~Kang, V.~Kansal, D.~Karagiannis, K.~Kiiveri, C.~C. Kirkpatrick, S.~Kruk, J.~Le Graet, L.~Legrand, M.~Lembo, F.~Lepori, G.~Leroy, G.~F. Lesci, J.~Lesgourgues, L.~Leuzzi, T.~I. Liaudat, A.~Loureiro, J.~Macias-Perez, G.~Maggio, M.~Magliocchetti, E.~A. Magnier, F.~Mannucci, R.~Maoli, C.~J. A.~P. Martins, L.~Maurin, M.~Miluzio, P.~Monaco, C.~Moretti, G.~Morgante, C.~Murray, K.~Naidoo, A.~Navarro-Alsina, S.~Nesseris, F.~Passalacqua, K.~Paterson, L.~Patrizii, A.~Pisani, D.~Potter, S.~Quai, M.~Radovich, S.~Sacquegna, M.~Sahlén, D.~B. Sanders, E.~Sarpa, A.~Schneider, D.~Sciotti, D.~Scognamiglio,
  E.~Sellentin, L.~C. Smith, K.~Tanidis, G.~Testera, R.~Teyssier, S.~Tosi, A.~Troja, M.~Tucci, C.~Valieri, A.~Venhola, D.~Vergani, G.~Verza, P.~Vielzeuf, N.~A. Walton, and J.~G. Sorce.
\newblock Euclid {Quick} {Data} {Release} ({Q1}) {Exploring} galaxy properties with a multi-modal foundation model, March 2025{\natexlab{b}}.
\newblock URL \url{http://arxiv.org/abs/2503.15312}.
\newblock arXiv:2503.15312 [astro-ph].

\bibitem[{Euclid Collaboration} et~al.(2025{\natexlab{c}}){Euclid Collaboration}, Walmsley, Huertas-Company, Quilley, Masters, Kruk, Remmelgas, Popp, Romelli, O'Ryan, Dickinson, Lintott, Serjeant, Smethurst, Simmons, Makechemu, Garland, Roberts, Mantha, Fortson, Géron, Keel, Baeten, Macmillan, Bovy, Casas, Leo, Sánchez, Katona, Kovács, Aghanim, Altieri, Amara, Andreon, Auricchio, Aussel, Baccigalupi, Baldi, Balestra, Bardelli, Basset, Battaglia, Bender, Biviano, Bonchi, Branchini, Brescia, Brinchmann, Camera, Cañas-Herrera, Capobianco, Carbone, Carretero, Castander, Castellano, Castignani, Cavuoti, Chambers, Cimatti, Colodro-Conde, Congedo, Conselice, Conversi, Copin, Courbin, Courtois, Cropper, Silva, Degaudenzi, Lucia, Giorgio, Dolding, Dole, Dubath, Duncan, Dupac, Dusini, Ealet, Escoffier, Fabricius, Farina, Farinelli, Faustini, Finelli, Fosalba, Fotopoulou, Frailis, Franceschi, Galeotta, George, Gillis, Giocoli, Gómez-Alvarez, Gracia-Carpio, Granett, Grazian, Grupp, Gwyn, Haugan, Hoekstra, Holmes,
  Hook, Hormuth, Hornstrup, Hudelot, Jahnke, Jhabvala, Joachimi, Keihänen, Kermiche, Kiessling, Kohley, Kubik, Kuijken, Kümmel, Kunz, Kurki-Suonio, Lahav, Boulc'h, Brun, Mignant, Liebing, Ligori, Lilje, Lindholm, Lloro, Mainetti, Maino, Maiorano, Mansutti, Marcin, Marggraf, Martinelli, Martinet, Marulli, Massey, Maurogordato, McCracken, Medinaceli, Mei, Melchior, Mellier, Meneghetti, Merlin, Meylan, Mora, Moresco, Moscardini, Nakajima, Neissner, Nichol, Niemi, Nightingale, Padilla, Paltani, Pasian, Pedersen, Percival, Pettorino, Pires, Polenta, Poncet, Popa, Pozzetti, Raison, Rebolo, Renzi, Rhodes, Riccio, Roncarelli, Rusholme, Saglia, Sakr, Sánchez, Sapone, Sartoris, Schewtschenko, Schneider, Schrabback, Scodeggio, Secroun, Seidel, Seiffert, Serrano, Simon, Sirignano, Sirri, Stanco, Steinwagner, Tallada-Crespí, Tavagnacco, Taylor, Teplitz, Tereno, Tessore, Toft, Toledo-Moreo, Torradeflot, Tutusaus, Valentijn, Valenziano, Valiviita, Vassallo, Kleijn, Veropalumbo, Wang, Weller, Zacchei, Zamorani, Zerbi,
  Zinchenko, Zucca, Allevato, Ballardini, Bolzonella, Bozzo, Burigana, Cabanac, Cappi, Ferdinando, Vigo, Gabarra, Martín-Fleitas, Matthew, Mauri, Metcalf, Pezzotta, Pöntinen, Porciani, Risso, Scottez, Sereno, Tenti, Viel, Wiesmann, Akrami, Andika, Anselmi, Archidiacono, Atrio-Barandela, Benoist, Benson, Bertacca, Bethermin, Bisigello, Blanchard, Blot, Böhringer, Brown, Bruton, Buitrago, Calabro, Quevedo, Caro, Carvalho, Castro, Cogato, Cooray, Cucciati, Davini, Paolis, Desprez, Díaz-Sánchez, Diaz, Domizio, Diego, Duc, Enia, Fang, Ferrari, Finoguenov, Fontana, Franco, Ganga, García-Bellido, Gasparetto, Gautard, Gaztanaga, Giacomini, Gozaliasl, Guidi, Gutierrez, Hall, Hartley, Hemmati, Hernández-Monteagudo, Hildebrandt, Hjorth, Kajava, Kang, Kansal, Karagiannis, Kiiveri, Kirkpatrick, Graet, Legrand, Lembo, Lepori, Leroy, Lesci, Lesgourgues, Leuzzi, Liaudat, Loureiro, Macias-Perez, Maggio, Magliocchetti, Mannucci, Maoli, Martins, Maurin, Miluzio, Monaco, Moretti, Morgante, Murray, Nadathur, Naidoo,
  Navarro-Alsina, Nesseris, Passalacqua, Paterson, Patrizii, Pisani, Potter, Quai, Radovich, Rocci, Rodighiero, Sacquegna, Sahlén, Sanders, Sarpa, Scarlata, Schaye, Schneider, Schultheis, Sciotti, Sellentin, Shankar, Smith, Tanidis, Testera, Teyssier, Tosi, Troja, Tucci, Valieri, Venhola, Vergani, Verza, Vielzeuf, Walton, Soubrie, and Scott]{euclidcollaborationEuclidQuickData2025b}
{}~{Euclid Collaboration}, M.~Walmsley, M.~Huertas-Company, L.~Quilley, K.~L. Masters, S.~Kruk, K.~A. Remmelgas, J.~J. Popp, E.~Romelli, D.~O'Ryan, H.~J. Dickinson, C.~J. Lintott, S.~Serjeant, R.~J. Smethurst, B.~Simmons, J.~Shingirai Makechemu, I.~L. Garland, H.~Roberts, K.~Mantha, L.~F. Fortson, T.~Géron, W.~Keel, E.~M. Baeten, C.~Macmillan, J.~Bovy, S.~Casas, C.~De Leo, H.~Domínguez Sánchez, J.~Katona, A.~Kovács, N.~Aghanim, B.~Altieri, A.~Amara, S.~Andreon, N.~Auricchio, H.~Aussel, C.~Baccigalupi, M.~Baldi, A.~Balestra, S.~Bardelli, A.~Basset, P.~Battaglia, R.~Bender, A.~Biviano, A.~Bonchi, E.~Branchini, M.~Brescia, J.~Brinchmann, S.~Camera, G.~Cañas-Herrera, V.~Capobianco, C.~Carbone, J.~Carretero, F.~J. Castander, M.~Castellano, G.~Castignani, S.~Cavuoti, K.~C. Chambers, A.~Cimatti, C.~Colodro-Conde, G.~Congedo, C.~J. Conselice, L.~Conversi, Y.~Copin, F.~Courbin, H.~M. Courtois, M.~Cropper, A.~Da Silva, H.~Degaudenzi, G.~De Lucia, A.~M.~Di Giorgio, C.~Dolding, H.~Dole, F.~Dubath, C.~A.~J. Duncan,
  X.~Dupac, S.~Dusini, A.~Ealet, S.~Escoffier, M.~Fabricius, M.~Farina, R.~Farinelli, F.~Faustini, F.~Finelli, P.~Fosalba, S.~Fotopoulou, M.~Frailis, E.~Franceschi, S.~Galeotta, K.~George, B.~Gillis, C.~Giocoli, P.~Gómez-Alvarez, J.~Gracia-Carpio, B.~R. Granett, A.~Grazian, F.~Grupp, S.~Gwyn, S.~V.~H. Haugan, H.~Hoekstra, W.~Holmes, I.~M. Hook, F.~Hormuth, A.~Hornstrup, P.~Hudelot, K.~Jahnke, M.~Jhabvala, B.~Joachimi, E.~Keihänen, S.~Kermiche, A.~Kiessling, R.~Kohley, B.~Kubik, K.~Kuijken, M.~Kümmel, M.~Kunz, H.~Kurki-Suonio, O.~Lahav, Q.~Le Boulc'h, A.~M. C.~Le Brun, D.~Le Mignant, P.~Liebing, S.~Ligori, P.~B. Lilje, V.~Lindholm, I.~Lloro, G.~Mainetti, D.~Maino, E.~Maiorano, O.~Mansutti, S.~Marcin, O.~Marggraf, M.~Martinelli, N.~Martinet, F.~Marulli, R.~Massey, S.~Maurogordato, H.~J. McCracken, E.~Medinaceli, S.~Mei, M.~Melchior, Y.~Mellier, M.~Meneghetti, E.~Merlin, G.~Meylan, A.~Mora, M.~Moresco, L.~Moscardini, R.~Nakajima, C.~Neissner, R.~C. Nichol, S.-M. Niemi, J.~W. Nightingale, C.~Padilla,
  S.~Paltani, F.~Pasian, K.~Pedersen, W.~J. Percival, V.~Pettorino, S.~Pires, G.~Polenta, M.~Poncet, L.~A. Popa, L.~Pozzetti, F.~Raison, R.~Rebolo, A.~Renzi, J.~Rhodes, G.~Riccio, M.~Roncarelli, B.~Rusholme, R.~Saglia, Z.~Sakr, A.~G. Sánchez, D.~Sapone, B.~Sartoris, J.~A. Schewtschenko, P.~Schneider, T.~Schrabback, M.~Scodeggio, A.~Secroun, G.~Seidel, M.~Seiffert, S.~Serrano, P.~Simon, C.~Sirignano, G.~Sirri, L.~Stanco, J.~Steinwagner, P.~Tallada-Crespí, D.~Tavagnacco, A.~N. Taylor, H.~I. Teplitz, I.~Tereno, N.~Tessore, S.~Toft, R.~Toledo-Moreo, F.~Torradeflot, I.~Tutusaus, E.~A. Valentijn, L.~Valenziano, J.~Valiviita, T.~Vassallo, G.~Verdoes Kleijn, A.~Veropalumbo, Y.~Wang, J.~Weller, A.~Zacchei, G.~Zamorani, F.~M. Zerbi, I.~A. Zinchenko, E.~Zucca, V.~Allevato, M.~Ballardini, M.~Bolzonella, E.~Bozzo, C.~Burigana, R.~Cabanac, A.~Cappi, D.~Di Ferdinando, J.~A.~Escartin Vigo, L.~Gabarra, J.~Martín-Fleitas, S.~Matthew, N.~Mauri, R.~B. Metcalf, A.~Pezzotta, M.~Pöntinen, C.~Porciani, I.~Risso, V.~Scottez,
  M.~Sereno, M.~Tenti, M.~Viel, M.~Wiesmann, Y.~Akrami, I.~T. Andika, S.~Anselmi, M.~Archidiacono, F.~Atrio-Barandela, C.~Benoist, K.~Benson, D.~Bertacca, M.~Bethermin, L.~Bisigello, A.~Blanchard, L.~Blot, H.~Böhringer, M.~L. Brown, S.~Bruton, F.~Buitrago, A.~Calabro, B.~Camacho Quevedo, F.~Caro, C.~S. Carvalho, T.~Castro, F.~Cogato, A.~R. Cooray, O.~Cucciati, S.~Davini, F.~De Paolis, G.~Desprez, A.~Díaz-Sánchez, J.~J. Diaz, S.~Di Domizio, J.~M. Diego, P.-A. Duc, A.~Enia, Y.~Fang, A.~G. Ferrari, A.~Finoguenov, A.~Fontana, A.~Franco, K.~Ganga, J.~García-Bellido, T.~Gasparetto, V.~Gautard, E.~Gaztanaga, F.~Giacomini, G.~Gozaliasl, M.~Guidi, C.~M. Gutierrez, A.~Hall, W.~G. Hartley, S.~Hemmati, C.~Hernández-Monteagudo, H.~Hildebrandt, J.~Hjorth, J.~J.~E. Kajava, Y.~Kang, V.~Kansal, D.~Karagiannis, K.~Kiiveri, C.~C. Kirkpatrick, J.~Le Graet, L.~Legrand, M.~Lembo, F.~Lepori, G.~Leroy, G.~F. Lesci, J.~Lesgourgues, L.~Leuzzi, T.~I. Liaudat, A.~Loureiro, J.~Macias-Perez, G.~Maggio, M.~Magliocchetti, F.~Mannucci,
  R.~Maoli, C.~J. A.~P. Martins, L.~Maurin, M.~Miluzio, P.~Monaco, C.~Moretti, G.~Morgante, C.~Murray, S.~Nadathur, K.~Naidoo, A.~Navarro-Alsina, S.~Nesseris, F.~Passalacqua, K.~Paterson, L.~Patrizii, A.~Pisani, D.~Potter, S.~Quai, M.~Radovich, P.-F. Rocci, G.~Rodighiero, S.~Sacquegna, M.~Sahlén, D.~B. Sanders, E.~Sarpa, C.~Scarlata, J.~Schaye, A.~Schneider, M.~Schultheis, D.~Sciotti, E.~Sellentin, F.~Shankar, L.~C. Smith, K.~Tanidis, G.~Testera, R.~Teyssier, S.~Tosi, A.~Troja, M.~Tucci, C.~Valieri, A.~Venhola, D.~Vergani, G.~Verza, P.~Vielzeuf, N.~A. Walton, E.~Soubrie, and D.~Scott.
\newblock Euclid {Quick} {Data} {Release} ({Q1}): {First} visual morphology catalogue, March 2025{\natexlab{c}}.
\newblock URL \url{http://arxiv.org/abs/2503.15310}.
\newblock arXiv:2503.15310 [astro-ph].

\bibitem[Fathkouhi and Fox(2024)]{fathkouhiAstroMAERedshiftPrediction2024}
Amirreza~Dolatpour Fathkouhi and Geoffrey~Charles Fox.
\newblock {AstroMAE}: {Redshift} {Prediction} {Using} a {Masked} {Autoencoder} with a {Novel} {Fine}-{Tuning} {Architecture}, September 2024.
\newblock URL \url{http://arxiv.org/abs/2409.01825}.
\newblock arXiv:2409.01825 [cs].

\bibitem[Feichtenhofer et~al.(2022)Feichtenhofer, Fan, Li, and He]{feichtenhoferMaskedAutoencodersSpatiotemporal2022}
Christoph Feichtenhofer, Haoqi Fan, Yanghao Li, and Kaiming He.
\newblock Masked {Autoencoders} {As} {Spatiotemporal} {Learners}, October 2022.
\newblock URL \url{http://arxiv.org/abs/2205.09113}.
\newblock arXiv:2205.09113 [cs].

\bibitem[Gao et~al.(2024)Gao, la~Tour, Tillman, Goh, Troll, Radford, Sutskever, Leike, and Wu]{gao2024scalingevaluatingsparseautoencoders}
Leo Gao, Tom~Dupré la~Tour, Henk Tillman, Gabriel Goh, Rajan Troll, Alec Radford, Ilya Sutskever, Jan Leike, and Jeffrey Wu.
\newblock Scaling and evaluating sparse autoencoders, 2024.
\newblock URL \url{https://arxiv.org/abs/2406.04093}.

\bibitem[Geirhos et~al.(2020)Geirhos, Jacobsen, Michaelis, Zemel, Brendel, Bethge, and Wichmann]{geirhosShortcutLearningDeep2020}
Robert Geirhos, Jörn-Henrik Jacobsen, Claudio Michaelis, Richard Zemel, Wieland Brendel, Matthias Bethge, and Felix~A. Wichmann.
\newblock Shortcut {Learning} in {Deep} {Neural} {Networks}.
\newblock \emph{Nature Machine Intelligence}, 2\penalty0 (11):\penalty0 665--673, November 2020.
\newblock ISSN 2522-5839.
\newblock \doi{10.1038/s42256-020-00257-z}.
\newblock URL \url{http://arxiv.org/abs/2004.07780}.
\newblock arXiv:2004.07780 [cs].

\bibitem[Grigg et~al.(2021)Grigg, Busbridge, Ramapuram, and Webb]{grigg2021selfsupervisedsupervisedmethodslearn}
Tom~George Grigg, Dan Busbridge, Jason Ramapuram, and Russ Webb.
\newblock Do self-supervised and supervised methods learn similar visual representations?, 2021.
\newblock URL \url{https://arxiv.org/abs/2110.00528}.

\bibitem[He et~al.(2022)He, Chen, Xie, Li, Dollár, and Girshick]{heMaskedAutoencodersAre2022}
Kaiming He, Xinlei Chen, Saining Xie, Yanghao Li, Piotr Dollár, and Ross Girshick.
\newblock Masked {Autoencoders} {Are} {Scalable} {Vision} {Learners}.
\newblock In \emph{2022 {IEEE}/{CVF} {Conference} on {Computer} {Vision} and {Pattern} {Recognition} ({CVPR})}, pages 15979--15988, June 2022.
\newblock \doi{10.1109/CVPR52688.2022.01553}.
\newblock URL \url{https://ieeexplore.ieee.org/document/9879206}.
\newblock ISSN: 2575-7075.

\bibitem[Karvonen et~al.(2025)Karvonen, Rager, Lin, Tigges, Bloom, Chanin, Lau, Farrell, McDougall, Ayonrinde, Till, Wearden, Conmy, Marks, and Nanda]{karvonenSAEBenchComprehensiveBenchmark2025}
Adam Karvonen, Can Rager, Johnny Lin, Curt Tigges, Joseph Bloom, David Chanin, Yeu-Tong Lau, Eoin Farrell, Callum McDougall, Kola Ayonrinde, Demian Till, Matthew Wearden, Arthur Conmy, Samuel Marks, and Neel Nanda.
\newblock {SAEBench}: {A} {Comprehensive} {Benchmark} for {Sparse} {Autoencoders} in {Language} {Model} {Interpretability}, June 2025.
\newblock URL \url{http://arxiv.org/abs/2503.09532}.
\newblock arXiv:2503.09532 [cs].

\bibitem[{Lochner} and {Bassett}(2021)]{2021A&C....3600481L}
M.~{Lochner} and B.~A. {Bassett}.
\newblock {ASTRONOMALY: Personalised active anomaly detection in astronomical data}.
\newblock \emph{Astronomy and Computing}, 36:\penalty0 100481, July 2021.
\newblock \doi{10.1016/j.ascom.2021.100481}.

\bibitem[Makhzani and Frey(2014)]{makhzani2014ksparseautoencoders}
Alireza Makhzani and Brendan Frey.
\newblock k-sparse autoencoders, 2014.
\newblock URL \url{https://arxiv.org/abs/1312.5663}.

\bibitem[Marks et~al.(2025)Marks, Treutlein, Bricken, Lindsey, Marcus, Mishra-Sharma, Ziegler, Ameisen, Batson, Belonax, Bowman, Carter, Chen, Cunningham, Denison, Dietz, Golechha, Khan, Kirchner, Leike, Meek, Nishimura-Gasparian, Ong, Olah, Pearce, Roger, Salle, Shih, Tong, Thomas, Rivoire, Jermyn, MacDiarmid, Henighan, and Hubinger]{marks2025auditinglanguagemodelshidden}
Samuel Marks, Johannes Treutlein, Trenton Bricken, Jack Lindsey, Jonathan Marcus, Siddharth Mishra-Sharma, Daniel Ziegler, Emmanuel Ameisen, Joshua Batson, Tim Belonax, Samuel~R. Bowman, Shan Carter, Brian Chen, Hoagy Cunningham, Carson Denison, Florian Dietz, Satvik Golechha, Akbir Khan, Jan Kirchner, Jan Leike, Austin Meek, Kei Nishimura-Gasparian, Euan Ong, Christopher Olah, Adam Pearce, Fabien Roger, Jeanne Salle, Andy Shih, Meg Tong, Drake Thomas, Kelley Rivoire, Adam Jermyn, Monte MacDiarmid, Tom Henighan, and Evan Hubinger.
\newblock Auditing language models for hidden objectives, 2025.
\newblock URL \url{https://arxiv.org/abs/2503.10965}.

\bibitem[Movva et~al.(2025)Movva, Peng, Garg, Kleinberg, and Pierson]{movva2025sparseautoencodershypothesisgeneration}
Rajiv Movva, Kenny Peng, Nikhil Garg, Jon Kleinberg, and Emma Pierson.
\newblock Sparse autoencoders for hypothesis generation, 2025.
\newblock URL \url{https://arxiv.org/abs/2502.04382}.

\bibitem[Muhamed et~al.(2024)Muhamed, Diab, and Smith]{muhamed2024decodingdarkmatterspecialized}
Aashiq Muhamed, Mona Diab, and Virginia Smith.
\newblock Decoding dark matter: Specialized sparse autoencoders for interpreting rare concepts in foundation models, 2024.
\newblock URL \url{https://arxiv.org/abs/2411.00743}.

\bibitem[{O'Neill} et~al.(2024){O'Neill}, {Ye}, {Iyer}, and {Wu}]{2024arXiv240800657O}
Charles {O'Neill}, Christine {Ye}, Kartheik {Iyer}, and John~F. {Wu}.
\newblock {Disentangling Dense Embeddings with Sparse Autoencoders}.
\newblock \emph{arXiv e-prints}, art. arXiv:2408.00657, August 2024.
\newblock \doi{10.48550/arXiv.2408.00657}.

\bibitem[O'Ryan and Gómez(2025)]{oryan2025identifyingastrophysicalanomalies996}
David O'Ryan and Pablo Gómez.
\newblock Identifying astrophysical anomalies in 99.6 million cutouts from the hubble legacy archive using anomalymatch, 2025.
\newblock URL \url{https://arxiv.org/abs/2505.03508}.

\bibitem[Peng et~al.(2025)Peng, Movva, Kleinberg, Pierson, and Garg]{peng2025usesparseautoencodersdiscover}
Kenny Peng, Rajiv Movva, Jon Kleinberg, Emma Pierson, and Nikhil Garg.
\newblock Use sparse autoencoders to discover unknown concepts, not to act on known concepts, 2025.
\newblock URL \url{https://arxiv.org/abs/2506.23845}.

\bibitem[Reed et~al.(2023)Reed, Gupta, Li, Brockman, Funk, Clipp, Keutzer, Candido, Uyttendaele, and Darrell]{reedScaleMAEScaleAwareMasked2023}
Colorado~J. Reed, Ritwik Gupta, Shufan Li, Sarah Brockman, Christopher Funk, Brian Clipp, Kurt Keutzer, Salvatore Candido, Matt Uyttendaele, and Trevor Darrell.
\newblock Scale-{MAE}: {A} {Scale}-{Aware} {Masked} {Autoencoder} for {Multiscale} {Geospatial} {Representation} {Learning}, September 2023.
\newblock URL \url{http://arxiv.org/abs/2212.14532}.
\newblock arXiv:2212.14532 [cs].

\bibitem[Scaramella et~al.(2021)Scaramella, Amiaux, Mellier, Burigana, Carvalho, Cuillandre, Silva, Derosa, Dinis, Maiorano, Maris, Tereno, Laureijs, Boenke, Buenadicha, Dupac, Venancio, Gómez-Álvarez, Hoar, Alvarez, Racca, Saavedra-Criado, Schwartz, Vavrek, Schirmer, Aussel, Azzollini, Cardone, Cropper, Ealet, Garilli, Gillard, Granett, Guzzo, Hoekstra, Jahnke, Kitching, Meneghetti, Miller, Nakajima, Niemi, Pasian, Percival, Sauvage, Scodeggio, Wachter, Zacchei, Aghanim, Amara, Auphan, Auricchio, Awan, Balestra, Bender, Bodendorf, Bonino, Branchini, Brau-Nogue, Brescia, Candini, Capobianco, Carbone, Carlberg, Carretero, Casas, Castander, Castellano, Cavuoti, Cimatti, Cledassou, Congedo, Conselice, Conversi, Copin, Corcione, Costille, Courbin, Degaudenzi, Douspis, Dubath, Duncan, Dusini, Farrens, Ferriol, Fosalba, Fourmanoit, Frailis, Franceschi, Franzetti, Fumana, Gillis, Giocoli, Grazian, Grupp, Haugan, Holmes, Hormuth, Hudelot, Kermiche, Kiessling, Kilbinger, Kohley, Kubik, Kümmel, Kunz, Kurki-Suonio,
  Ligori, Lilje, Lloro, Mansutti, Marggraf, Markovic, Marulli, Massey, Maurogordato, Melchior, Merlin, Meylan, Mohr, Moresco, Morin, Moscardini, Munari, Nichol, Padilla, Paltani, Peacock, Pedersen, Pettorino, Pires, Poncet, Popa, Pozzetti, Raison, Rebolo, Rhodes, Rix, Roncarelli, Rossetti, Saglia, Schneider, Schrabback, Secroun, Seidel, Serrano, Sirignano, Sirri, Skottfelt, Stanco, Starck, Tallada-Crespí, Tavagnacco, Taylor, Teplitz, Toledo-Moreo, Torradeflot, Trifoglio, Valentijn, Valenziano, Kleijn, Wang, Welikala, Weller, Wetzstein, Zamorani, Zoubian, Andreon, Baldi, Bardelli, Boucaud, Camera, Fabbian, Farinelli, Graciá-Carpio, Maino, Medinaceli, Mei, Neissner, Polenta, Renzi, Romelli, Rosset, Sureau, Tenti, Vassallo, Zucca, Baccigalupi, Balaguera-Antolínez, Battaglia, Biviano, Borgani, Bozzo, Cabanac, Cappi, Casas, Castignani, Colodro-Conde, Coupon, Courtois, Cuby, Torre, Desai, Ferdinando, Dole, Fabricius, Farina, Ferreira, Finelli, Flose-Reimberg, Fotopoulou, Galeotta, Ganga, Gozaliasl, Hook,
  Keihanen, Kirkpatrick, Liebing, Lindholm, Mainetti, Martinelli, Martinet, Maturi, McCracken, Metcalf, Morgante, Nightingale, Nucita, Patrizii, Potter, Riccio, Sánchez, Sapone, Schewtschenko, Schultheis, Scottez, Teyssier, Tutusaus, Valiviita, Viel, Vriend, and Whittaker]{scaramellaEuclidPreparationEuclid2021}
R.~Scaramella, J.~Amiaux, Y.~Mellier, C.~Burigana, C.~S. Carvalho, J.-C. Cuillandre, A.~Da Silva, A.~Derosa, J.~Dinis, E.~Maiorano, M.~Maris, I.~Tereno, R.~Laureijs, T.~Boenke, G.~Buenadicha, X.~Dupac, L.~M.~Gaspar Venancio, P.~Gómez-Álvarez, J.~Hoar, J.~Lorenzo Alvarez, G.~D. Racca, G.~Saavedra-Criado, J.~Schwartz, R.~Vavrek, M.~Schirmer, H.~Aussel, R.~Azzollini, V.~F. Cardone, M.~Cropper, A.~Ealet, B.~Garilli, W.~Gillard, B.~R. Granett, L.~Guzzo, H.~Hoekstra, K.~Jahnke, T.~Kitching, M.~Meneghetti, L.~Miller, R.~Nakajima, S.~M. Niemi, F.~Pasian, W.~J. Percival, M.~Sauvage, M.~Scodeggio, S.~Wachter, A.~Zacchei, N.~Aghanim, A.~Amara, T.~Auphan, N.~Auricchio, S.~Awan, A.~Balestra, R.~Bender, C.~Bodendorf, D.~Bonino, E.~Branchini, S.~Brau-Nogue, M.~Brescia, G.~P. Candini, V.~Capobianco, C.~Carbone, R.~G. Carlberg, J.~Carretero, R.~Casas, F.~J. Castander, M.~Castellano, S.~Cavuoti, A.~Cimatti, R.~Cledassou, G.~Congedo, C.~J. Conselice, L.~Conversi, Y.~Copin, L.~Corcione, A.~Costille, F.~Courbin, H.~Degaudenzi,
  M.~Douspis, F.~Dubath, C.~A.~J. Duncan, S.~Dusini, S.~Farrens, S.~Ferriol, P.~Fosalba, N.~Fourmanoit, M.~Frailis, E.~Franceschi, P.~Franzetti, M.~Fumana, B.~Gillis, C.~Giocoli, A.~Grazian, F.~Grupp, S.~V.~H. Haugan, W.~Holmes, F.~Hormuth, P.~Hudelot, S.~Kermiche, A.~Kiessling, M.~Kilbinger, R.~Kohley, B.~Kubik, M.~Kümmel, M.~Kunz, H.~Kurki-Suonio, S.~Ligori, P.~B. Lilje, I.~Lloro, O.~Mansutti, O.~Marggraf, K.~Markovic, F.~Marulli, R.~Massey, S.~Maurogordato, M.~Melchior, E.~Merlin, G.~Meylan, J.~J. Mohr, M.~Moresco, B.~Morin, L.~Moscardini, E.~Munari, R.~C. Nichol, C.~Padilla, S.~Paltani, J.~Peacock, K.~Pedersen, V.~Pettorino, S.~Pires, M.~Poncet, L.~Popa, L.~Pozzetti, F.~Raison, R.~Rebolo, J.~Rhodes, H.-W. Rix, M.~Roncarelli, E.~Rossetti, R.~Saglia, P.~Schneider, T.~Schrabback, A.~Secroun, G.~Seidel, S.~Serrano, C.~Sirignano, G.~Sirri, J.~Skottfelt, L.~Stanco, J.~L. Starck, P.~Tallada-Crespí, D.~Tavagnacco, A.~N. Taylor, H.~I. Teplitz, R.~Toledo-Moreo, F.~Torradeflot, M.~Trifoglio, E.~A. Valentijn,
  L.~Valenziano, G.~A.~Verdoes Kleijn, Y.~Wang, N.~Welikala, J.~Weller, M.~Wetzstein, G.~Zamorani, J.~Zoubian, S.~Andreon, M.~Baldi, S.~Bardelli, A.~Boucaud, S.~Camera, G.~Fabbian, R.~Farinelli, J.~Graciá-Carpio, D.~Maino, E.~Medinaceli, S.~Mei, C.~Neissner, G.~Polenta, A.~Renzi, E.~Romelli, C.~Rosset, F.~Sureau, M.~Tenti, T.~Vassallo, E.~Zucca, C.~Baccigalupi, A.~Balaguera-Antolínez, P.~Battaglia, A.~Biviano, S.~Borgani, E.~Bozzo, R.~Cabanac, A.~Cappi, S.~Casas, G.~Castignani, C.~Colodro-Conde, J.~Coupon, H.~M. Courtois, J.~Cuby, S.~de~la Torre, S.~Desai, D.~Di Ferdinando, H.~Dole, M.~Fabricius, M.~Farina, P.~G. Ferreira, F.~Finelli, P.~Flose-Reimberg, S.~Fotopoulou, S.~Galeotta, K.~Ganga, G.~Gozaliasl, I.~M. Hook, E.~Keihanen, C.~C. Kirkpatrick, P.~Liebing, V.~Lindholm, G.~Mainetti, M.~Martinelli, N.~Martinet, M.~Maturi, H.~J. McCracken, R.~B. Metcalf, G.~Morgante, J.~Nightingale, A.~Nucita, L.~Patrizii, D.~Potter, G.~Riccio, A.~G. Sánchez, D.~Sapone, J.~A. Schewtschenko, M.~Schultheis, V.~Scottez,
  R.~Teyssier, I.~Tutusaus, J.~Valiviita, M.~Viel, W.~Vriend, and L.~Whittaker.
\newblock Euclid preparation: {I}. {The} {Euclid} {Wide} {Survey}, August 2021.
\newblock URL \url{http://arxiv.org/abs/2108.01201}.
\newblock arXiv:2108.01201.

\bibitem[Spergel et~al.(2015)Spergel, Gehrels, Baltay, Bennett, Breckinridge, Donahue, Dressler, Gaudi, Greene, Guyon, Hirata, Kalirai, Kasdin, Macintosh, Moos, Perlmutter, Postman, Rauscher, Rhodes, Wang, Weinberg, Benford, Hudson, Jeong, Mellier, Traub, Yamada, Capak, Colbert, Masters, Penny, Savransky, Stern, Zimmerman, Barry, Bartusek, Carpenter, Cheng, Content, Dekens, Demers, Grady, Jackson, Kuan, Kruk, Melton, Nemati, Parvin, Poberezhskiy, Peddie, Ruffa, Wallace, Whipple, Wollack, and Zhao]{spergel2015widefieldinfrarredsurveytelescopeastrophysics}
D.~Spergel, N.~Gehrels, C.~Baltay, D.~Bennett, J.~Breckinridge, M.~Donahue, A.~Dressler, B.~S. Gaudi, T.~Greene, O.~Guyon, C.~Hirata, J.~Kalirai, N.~J. Kasdin, B.~Macintosh, W.~Moos, S.~Perlmutter, M.~Postman, B.~Rauscher, J.~Rhodes, Y.~Wang, D.~Weinberg, D.~Benford, M.~Hudson, W.~S. Jeong, Y.~Mellier, W.~Traub, T.~Yamada, P.~Capak, J.~Colbert, D.~Masters, M.~Penny, D.~Savransky, D.~Stern, N.~Zimmerman, R.~Barry, L.~Bartusek, K.~Carpenter, E.~Cheng, D.~Content, F.~Dekens, R.~Demers, K.~Grady, C.~Jackson, G.~Kuan, J.~Kruk, M.~Melton, B.~Nemati, B.~Parvin, I.~Poberezhskiy, C.~Peddie, J.~Ruffa, J.~K. Wallace, A.~Whipple, E.~Wollack, and F.~Zhao.
\newblock Wide-field infrarred survey telescope-astrophysics focused telescope assets wfirst-afta 2015 report, 2015.
\newblock URL \url{https://arxiv.org/abs/1503.03757}.

\bibitem[{Wu}(2025)]{2025ApJ...980..183W}
John~F. {Wu}.
\newblock {Insights into Galaxy Evolution from Interpretable Sparse Feature Networks}.
\newblock \emph{The Astrophysical Journal}, 980\penalty0 (2):\penalty0 183, February 2025.
\newblock \doi{10.3847/1538-4357/adadec}.
\newblock URL \url{https://iopscience.iop.org/article/10.3847/1538-4357/adadec}.

\bibitem[Ye et~al.(2024)Ye, O'Neill, Wu, and Iyer]{ye2024steering}
Christine Ye, Charles O'Neill, John~F Wu, and Kartheik~G. Iyer.
\newblock Steering semantic search with interpretable features from sparse autoencoders.
\newblock In \emph{MINT: Foundation Model Interventions}, 2024.
\newblock URL \url{https://openreview.net/forum?id=oacksuh7Tu}.

\end{thebibliography}

\newpage
\appendix
\section{Interpreting SAE features beyond the top 64} \label{app:beyond}

\begin{figure}[ht]
    \centering
    \includegraphics[width=0.495\textwidth]{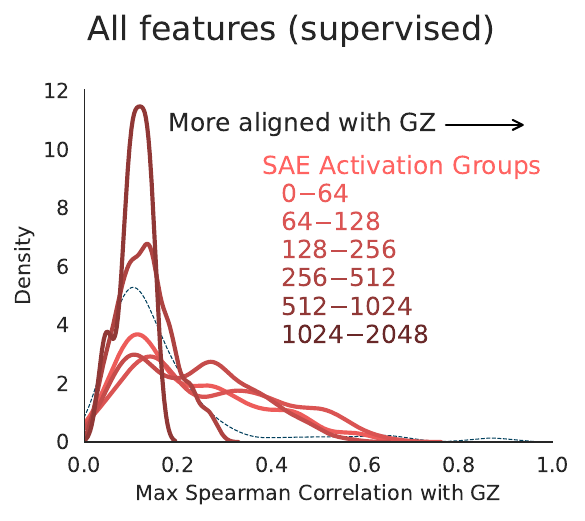}
    \includegraphics[width=0.495\textwidth]{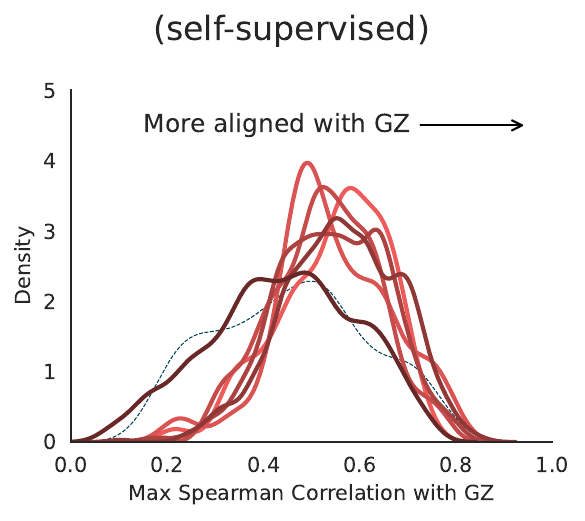}

    \caption{
    How aligned are features with GZ, just like in Figure~\ref{fig:feature_alignment}?
    We show results for both supervised (\textit{left}) and self-supervised (\textit{right}) approaches.
    This time we use all of the SAE activations, binned by their Matryoshka group sizes (progressing from lighter to darker red). 
    In all panels, the top 64 elements from PCA are shown in a thin dashed blue line.
    }
    \label{fig:appendix_feature_alignment}
\end{figure}

In Figure~\ref{fig:appendix_feature_alignment}, we show that SAE features beyond the top 64 are still aligned with GZ. For the supervised case (left-hand panel), the correlation grows weaker as we reach the later groups in the Matryoshka SAE: typical correlations for the top 64 features are $\sim 0.3$, whereas mean Spearman correlations in the final group with GZ are near $\sim 0.1$. 

Meanwhile, the SAE features for self-supervised embeddings (right-hand panel) all have a moderately high alignment with GZ features. Intriguingly, this does not vary much with the Matryoshka groups, suggesting that the basis set of MAE features is independent of GZ.

We also show PCA again, which exhibits significantly lower alignment with GZ than the SAE features in both the supervised and self-supervised cases first principal component, which has maximum $|r| \approx 0.87$ and separates smooth galaxies from disk or featured galaxies). Principal components beyond the top 64 have nearly zero max Spearman correlation $r$ with GZ features (and are consistent with random), so we omit them from Figure~\ref{fig:appendix_feature_alignment}.

\section{Feature Predictability} \label{app:predictability}
In the same spirit as the ``Feature Alignment'' analysis, we can also fit linear regression to predict each learned feature from all Galaxy Zoo labels and measure $R^2$. This is not a trivial deprojection for the supervised case because the GZ model has a non-linear classification head. 

The $R^2$ metric probes how \textit{predictable} any SAE feature is, based on known GZ features. If we are unable to explain the feature variance using existing morphology classifications, then it may suggest that the feature is novel. (It could also signify that the feature is highly polysemantic!) 

\begin{figure}[ht]
    \centering
    \includegraphics[width=0.495\textwidth]{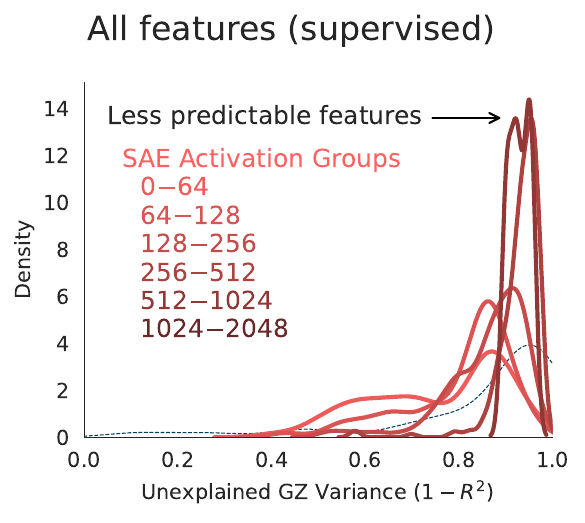}
    \includegraphics[width=0.495\textwidth]{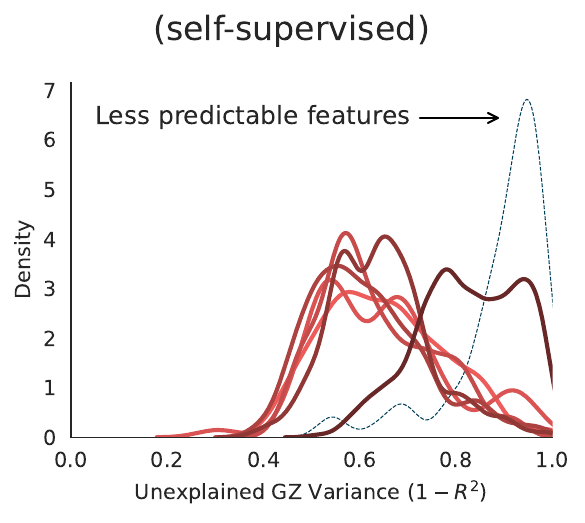}
    \caption{
     How unpredictable---or perhaps, ``novel''---are our features via linear combinations of GZ vote fractions?
     Figure in same style as Figure~\ref{fig:appendix_feature_alignment}.
    }     \label{fig:appendix_feature_novelty}
\end{figure}

In Figure~\ref{fig:appendix_feature_novelty}, we show how well GZ classes can predict the strengths of SAE activations and principal components. We find qualitatively similar results to before: for the supervised case, the PCA features are generally difficult to predict (aside from the first handful), and the SAE activations also become harder to predict as we progress to later Matryoshka groups. For the MAE embeddings, the SAE features are moderately predictable (and PCA features less so). Again, we suspect this is because MAE embeddings contain salient information about image artifacts, and GZ vote fractions can partially reconstruct these features.

\section{Expected Spearman Correlations with Random Features} \label{app:random-spearman}

We can fashion a null test by sampling random features and computing the expectation value of the maximum Spearman $r$ by using extreme value statistics:
$\mathbb{E}[{\rm max} (r)] \approx \sqrt{2 \log(2c)/(n-1)}$, where we have $c=46$ GZ classes, $k=64$ random features, and $n=2378$ or $76061$ test set samples respectively in the supervised and self-supervised test sets.
This approximation is valid in the limit of moderate to large $n$, and we use $2c$ rather than $c$ because we take the absolute Spearman correlation (probing a two-tailed distribution).
The standard deviation of the maximum Spearman rank can be approximated via $\pi / \sqrt{k(n-1)\log (2c)}$. These expectation values approximate to $0.041 \pm 0.002$ for the supervised case, and $0.0072 \pm 0.0003$ for self-supervised model embeddings.

\clearpage

\end{document}